\title{Aberrational Effects for Shadows of Black Holes}
\author{Arne Grenzebach\footnote{Email: arne.grenzebach@zarm.uni-bremen.de} \\
        ZARM\\
        University of Bremen\\
        Am Fallturm, 28359 Bremen, Germany}

\date{November 6, 2014}
\documentclass[11pt,english]{article}%,draft

\usepackage{babel}
\usepackage[T1]{fontenc}
\usepackage{amssymb}%,textcomp,mathcomp
\usepackage{amsmath}%[fleqn]
\usepackage{xcolor,graphicx,array}
\newcolumntype{C}{>{$}c<{$}}

\newcommand{\R}{\mathbb{R}}
\newcommand{\Lag}{\mathcal{L}}
\newcommand{\K}{\mathcal{K}}

\renewcommand{\vec}[1]{\boldsymbol{#1}}

\newcommand{\Menge}[3][big]{\csname#1l\endcsname\{#2\csname#1m\endcsname\vert#3\csname#1r\endcsname\}}
\newcommand{\abs}[1]{\lvert#1\rvert}

\newcommand{\unit}[1]{\:\mathrm{\textstyle #1}}

\let\blqq=\textquotedblleft
\let\brqq=\textquotedblright

\makeatletter
\newcommand{\drm}{\@ifstar{\mathrm{d}}{\:\mathrm{d}}}
\makeatother

\graphicspath{{Bilder/}}

\begin{document}
\maketitle

\begin{abstract}
In this paper, we discuss how the shadow of a Kerr black hole depends
on the motion of the observer. In particular, we derive an analytical 
formula for the boundary curve of the shadow for an observer moving 
with given four-velocity at given Boyer--Lindquist coordinates. We 
visualize the shadow for various values of parameters.
\end{abstract}

%%%%%%%%%%%%%%%%%%%%%%%%%%%%%%%%%%%%%%%%%%%%%%%%%%%%%%%%%%%%%%%%%%%%%%
\section{Introduction: What is the \blqq Shadow\brqq?}
\label{sec:Intro}
\nocite{GrenzebachPerlick.2014}
For an observer at Boyer--Lindquist coordinates $(r_O, \vartheta_O)$
the \emph{shadow} of a black hole is defined as that region of the 
observer's celestial sphere which is left dark if all light sources 
are distributed on a sphere with radius $r_L>r_O$.
To calculate the shadow, we consider light rays, i.e. lightlike 
geodesics, which are sent into the past from the observer's position. 
Then, the boundary curve of the shadow corresponds to the limiting 
case between geodesics going towards the horizon ($\rightarrow$ 
darkness) and geodesics going to the sphere at $r_L$ with light 
sources ($\rightarrow$ brightness). The limiting case are geodesics 
that asymptotically spiral towards (unstable) lightlike geodesics 
which fill a specific spatial region, the \emph{photon region} $\K$, 
and are propagating on a sphere. Consequently, the shadow of the black 
hole is a mapping of this photon region $\K$ and not of the horizon.

Starting with the theoretical work of Bardeen who for the first time 
calculated the shadow of a Kerr black hole correctly \cite{Bardeen.1973} 
we now reach the time where it seems to be possible to observe the 
shadow of the black hole near Sgr~A* at the center of our Galaxy. 
The expected image is shown by calculations of how the shadow of a 
black hole would look like if matter, in the form of an accretion 
disc, a corona, or a jet, is included in the model. These calculations 
are based on ray tracing and GRMHD, see e.g.
\cite{FalckeMelia.2000,DexterAgol.2012,DexterFragile.2013,
YounsiWu.2012,MoscibrodzkaFalcke.2014,FishJohnson.2014}. It is hoped that the 
shadow of Sgr~A* will indeed be observed in the near future within 
the \emph{Event Horizon Telescope} project, see \cite{Doeleman.2009}, or the 
\emph{Black Hole Cam} project.

%observations, Black Hole Cam (european project), Event Horizon Telescope
%theoretical: \cite{Bardeen.1973} \cite{Chandrasekhar.1983}
%GRMHD: 2 paper \cite{DexterFragile.2013} \cite{MoscibrodzkaFalcke.2014}
%observations: \cite{FalckeMelia.2000} \cite{DexterAgol.2012}

In the following, we discuss how differently moving observers at a given 
position see the shadow of a Kerr black hole. A detailed discussion of a 
more general space-time describing a Kerr--Newman--NUT black hole with 
cosmological constant can be found in \cite{GrenzebachPerlick.2014}. 
There, we demonstrate how the shadow is influenced by a charge, the 
cosmological constant or the NUT parameter. Since there one can also 
find a discussion of metric properties and visualizations of the photon 
regions of the black holes, we restrict ourselves in this proceedings 
volume to the Kerr case. 
Our construction of the shadow is a geometrical one, based on the 
geodesic equation and ignoring the influence of matter.

%%%%%%%%%%%%%%%%%%%%%%%%%%%%%%%%%%%%%%%%%%%%%%%%%%%%%%%%%%%%%%%%%%%%%%
\section{The Kerr Metric}
\label{sec:Metric}
The Kerr space-time is a stationary, axially symmetric type D solutions 
of the Einstein vacuum equations that describes a rotating black hole with 
mass $m$ and spin $a$. 
%It is a special case of the Pleba{\'n}ski--Demia{\'n}ski space-time which 
%provides additional parameters for electric and magnetic charge 
%($\beta=q_{e}^{2}+q_{m}^{2}$), for the gravitomagnetic NUT charge $\ell$, 
%for the cosmological constant $\Lambda$, and for an acceleration ($\alpha$). 
%The book of Griffiths and Podolsk{\'y} \cite{GriffithsPodolsky.2009} 
%includes a detailed discussion of the Pleba{\'n}ski\mbox{--De}mia{\'n}ski
%space-times; see also Stephani et al. \cite{StephaniKramer.2003}.
In Boyer--Lindquist coordinates $(r, \vartheta, \varphi, t)$ the 
Kerr metric can be written as (\cite{GriffithsPodolsky.2009}, p. 314)
%\citep[p. 314]{GriffithsPodolsky.2009}
\begin{equation}
\begin{aligned}
	g_{\mu \nu} \drm x^{\mu} \drm x^{\nu} &= 
		\Sigma \bigl( \tfrac{1}{\Delta}\drm r^2 + \drm \vartheta^2 \bigr)
	+ \tfrac{1}{\Sigma} 
		\bigl( (\Sigma + a\chi)^2 \sin^2\vartheta 
			- \Delta \chi^2 \bigr) \drm \varphi^2 \\
	&\quad + \tfrac{2}{\Sigma} 
		\bigl( \Delta\chi - a(\Sigma + a\chi) 
			\sin^2\vartheta \bigr) \drm t \drm \varphi
	- \tfrac{1}{\Sigma} 
		\bigl( \Delta - a^2\sin^2\vartheta \bigr) \drm t^2
\end{aligned}
	\label{eq:Metric}
\end{equation}
where we use the abbreviations
\begin{align}
	\Sigma &= r^2 + \bigl(a\cos\vartheta \bigr)^2, &
	\chi   &= a\sin^2\vartheta, &
	\Delta &= r^2 - 2mr + a^2,
	\label{eq:MetricFunc}
\end{align}
%\begin{equation}
%\begin{aligned}
%	\Sigma &= r^2 + \bigl(a\cos\vartheta \bigr)^2, \\
%	\chi   &= a\sin^2\vartheta, \\
%	\Delta &= r^2 - 2mr + a^2,
%\end{aligned}
%	\label{eq:MetricFunc}
%\end{equation}
and rescaled units; hence, the speed of light and the gravitational 
constant are normalized ($c=1$, $G=1$). 
The coordinates $t$ and $r$ range over $]{-}\infty, \infty[\,$, 
while $\vartheta$ and $\varphi$ are standard angular coordinates on
the two-sphere.  
Whereas the parameter $m$ for the mass of the black hole could take 
all values in $\R^+$ the absolute value of the spin parameter $a$ 
is bounded by $m$ since the event horizon of the Kerr black hole is 
at $r+\sqrt{m^2-a^2}$.

%%%%%%%%%%%%%%%%%%%%%%%%%%%%%%%%%%%%%%%%%%%%%%%%%%%%%%%%%%%%%%%%%%%%%%
\section{Calculating the Shadows of Black Holes}
\label{sec:Calc}
As the geodesic equation in the Kerr space-time has four 
constants of motion---the Lagrangian 
$\Lag$, the energy and the $z$-component of angular momentum 
\begin{align}
	E &= -\frac{\partial\Lag}{\partial\dot{t}} 
		= -g_{\varphi t}\dot{\varphi} -g_{tt}\dot{t}, \quad&\quad
	L_{z} &= \frac{\partial\Lag}{\partial\dot{\varphi}}
		= g_{\varphi\varphi}\dot{\varphi} +g_{\varphi t}\dot{t},
	\label{eq:ELz}
\end{align}
plus the Carter constant $K$ \cite{Carter.1968b}---the lightlike 
geodesics ($\Lag=0$) are given by four separated equations of motion
\begin{subequations}\label{eq:EoM}
	\begin{align}
	\Sigma \dot{t}
	&= \frac{\chi(L_{z}-E\chi)}{\sin^{2}\vartheta}
		+ \frac{(\Sigma + a\chi) \bigl((\Sigma + a\chi)E - aL_{z}\bigr)}{\Delta},
\label{eq:EoM_t} \\
	\Sigma \dot{\varphi}
	&= \frac{L_{z}-E\chi}{\sin^{2}\vartheta}
		+\frac{a\bigl((\Sigma + a\chi)E - aL_{z}\bigr)}{\Delta},
\label{eq:EoM_phi} \\
	\Sigma^{2} \dot{\vartheta}^{2} 
	&= K - \frac{(\chi E - L_{z})^{2}}{\sin^{2}\vartheta} 
	=: \Theta(\vartheta),
\label{eq:EoM_theta} \\
	\Sigma^{2} \dot{r}^{2} 
	&= \bigl((\Sigma + a\chi)E-aL_{z}\bigr)^{2} - \Delta K
	=: R(r).
\label{eq:EoM_r}
\end{align}
\end{subequations}

The existence of the \emph{photon region} $\K$, i.e. the region filled 
with lightlike geodesics staying on a sphere $r=\mathrm{constant}$, 
is crucial for calculating the shadow because it will give us a 
parametrization of the shadow's boundary curve.
By \eqref{eq:EoM_r}, the sphere conditions $\dot{r}=0$ and $\ddot{r}=0$ 
imply that $R(r)=0$ and $R'(r)=0$. Hence, 
\begin{align}
	K_{E} &= \frac{\bigl((\Sigma + a\chi)-aL_{E}\bigr)^{2}}{\Delta}, &
	K_{E} &= \frac{2r\bigl((\Sigma + a\chi)-aL_{E}\bigr)}{r-m},
\label{eq:2xKE}
\end{align}
where $L_{E}=\frac{L_{z}}{E}$ and $K_{E}=\frac{K}{E^{2}}$. Solving for 
these constants gives the expressions
\begin{align}
	K_{E} &= \frac{4r^{2}\Delta}{(r-m)^{2}}, &
	aL_{E} &= \bigl(\Sigma + a\chi\bigr) - \frac{2r\Delta}{r-m}
\label{eq:KL_SphLR}
\end{align}
for spherical light rays. Since 
$0\leq \Sigma^{2} \dot{\vartheta}^{2}$ wie find by \eqref{eq:EoM_theta} 
an inequality characterizing the photon region
\begin{equation}
	\K\colon \bigl(2r\Delta - \Sigma (r-m) \bigr)^{2}
		\leq 4 a^{2} r^{2} \Delta \sin^{2}\vartheta.
\label{eq:regionK}
\end{equation}
Through each point $(r,\vartheta)$ of $\K$ there is a light ray propagating 
on a sphere. Plots and a detailed discussion of the photon region for different 
space-times can be found in \cite[Fig.~3--5]{GrenzebachPerlick.2014}.

%%%%%%%%%%%%%%%%%%%%%%%%%%%%%%%%%%%%%%%%%%%%%%%%%%%%%%%%%%%%%%%%%%%%%%
\section{Viewing the Shadows of Black Holes}
\label{sec:View}

For determining the shadow of a black hole we consider an observer 
at Boyer-Lindquist coordinates $(r_{O}, \vartheta_{O})$ and assume, 
for the sake of simplicity, that the light sources are distributed 
on a sphere with radius $r_{L}>r_{O}$.

Lightlike geodesics reaching the observer can be divided into two 
types of orbits. There are geodesics which passed the sphere with 
light sources and there are those coming from the horizon. 
Thus, our observer would see brightness in the direction
of light rays of the first type and darkness for the other ones. 
The boundary curve of the shadow is therefore given by lightlike 
geodesics that spiraled from one of the unstable spherical light 
orbits of the photon region $\K$. 

The shape of the shadow depends on the observer's state of motion.
At Boyer--Lindquist coordinates $(r_{O},\vartheta_{O})$, we choose 
an orthonormal tetrad adapted to the symmetry of the space-time 
(\cite{GriffithsPodolsky.2009}, p.~307)
%\citep[p.~307]{GriffithsPodolsky.2009}
\begin{equation}
\begin{aligned}
	e_{0} &= \left.
		\frac{(\Sigma + a \chi) \partial_t + a \partial_{\varphi}}{
		\sqrt{\Sigma \Delta}}\right|_{(r_O,\vartheta_O)}, \quad &\quad
	e_{1} &= \left.
		\sqrt{\dfrac{1}{\Sigma}} \, \partial_{\vartheta}
		\right|_{(r_O,\vartheta_O)}, \\[\smallskipamount]
	e_{2} &= -\left.
		\frac{\partial_{\varphi} + \chi \partial_t}{
		\sqrt{\Sigma}  \sin \vartheta} 
		\right|_{(r_O,\vartheta_O)}, \quad &\quad
	e_{3} &= -\left.
		\sqrt{\frac{\Delta}{\Sigma}} \, \partial_r
		\right|_{(r_O,\vartheta_O)}.
\end{aligned}
\label{eq:ObsTetrad}
\end{equation}
It is chosen such that $e_0 \pm e_3$ are tangential to the \emph{principal 
null congruences} of our metric. Here, $e_0$ is interpreted as the 
four-velocity of an observer at $(r_{O},\vartheta_{O})$ because it is 
a timelike vector;  $e_3$ points into the direction towards the center of 
the black hole.
An observer with this tetrad is called a \emph{standard observer} in the 
following.

\begin{figure}[btp]
\centering
	\includegraphics{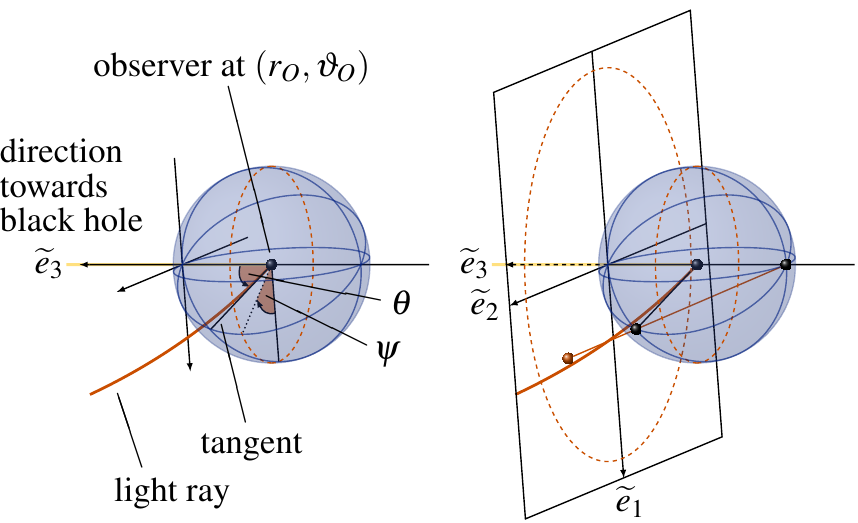}
	\caption{The direction of each light ray reaching the observer
	is given by the celestial coordinates 
	$\theta$ and $\psi$ (Eq.~\eqref{eq:dotlambda2a}) of their tangents, 
	see figure on the left. %
	These points $(\theta,\psi)$ on the celestial sphere (black ball)
	can be identified with points in the plane (red ball) by 
	stereographic projection, see figure on the right. 
	The dashed (red) circles mark the celestial equator 
	$\theta = \pi/2$ respectively its projection.}
	\label{fig:Beob}
\end{figure}

%%%%%%%%%%%%%%%%%%%%%%%%%%%%%%%%%%%%%%%%%%%%%%%%%%%%%%%%%%%%%%%%%%%%%%
If another observer at $(r_{O},\vartheta_{O})$ moves with velocity 
$\vec{v}=(v_{1},v_{2},v_{3})$, $v=\abs{\vec{v}}<1=c$, with respect 
to our standard observer, we have to modify the tetrad.
The four-velocity of the moving observer is 
\begin{subequations}\label{eq:ObsTetradv}
\begin{align}
	& \widetilde{e}_{0} = \frac{v_{1}e_{1} + v_{2}e_{2} + v_{3}e_{3} + e_{0}}{
		\sqrt{1-v^{2}}}.
	\label{eq:ObsTetradv0}
\intertext{From $\widetilde{e}_{0},e_{1},e_{2},e_{3}$ we find an 
orthonormal tetrad 
$\widetilde{e}_{0},\widetilde{e}_{1},\widetilde{e}_{2},\widetilde{e}_{3}$
with the Gram--Schmidt procedure by adding $e_{3}$, $e_{1}$, $e_{2}$--in 
this order--successively to $\widetilde{e}_{0}$}
	&\begin{aligned}
		\widetilde{e}_{1} &= \frac{\bigl(1-v_{2}^{2}\bigr)e_{1} 
			+ v_{1}(v_{2}e_{2}+e_{0})}{ \sqrt{1-v_{2}^{2}} \; 
			\sqrt{1-v_{1}^{2}-v_{2}^{2}} }, \\[\smallskipamount]
		\widetilde{e}_{2} &= \frac{ e_{2} + v_{2}e_{0} }{ \sqrt{1-v_{2}^{2}} }, 
			\\[\smallskipamount]
		\widetilde{e}_{3} &= \frac{\bigl(1-v_{1}^{2}-v_{2}^{2}\bigr)e_{3} 
			+ v_{3}(v_{1}e_{1}+v_{2}e_{2}+e_{0})}{ \sqrt{1-v_{1}^{2}-v_{2}^{2}} \; 				\sqrt{1-v^{2}} }.
	\end{aligned}
	\label{eq:ObsTetradvi}
\end{align}
\end{subequations}
Note that $\widetilde{e}_{i}=e_{i}$ if $v_{i}=0$, i.e., for $\vec{v}=0$ this 
procedure recovers the tetrad $e_{0}, e_{1}, e_{2}, e_{3}$. %
As before, the spacelike vector $\widetilde{e}_{3}$ corresponds to the 
direction towards the black hole. The interpretation of $\widetilde{e}_{1}$ and 
$\widetilde{e}_{2}$ becomes clear if we introduce celestial coordinates, see 
\eqref{eq:dotlambda2a} and Fig.~\ref{fig:Beob}. Then, 
% $\widetilde{e}_{3}$ is the pole of the celestial coordinates while 
$\widetilde{e}_{1}$ and $\widetilde{e}_{2}$ point into the 
north--south respectively the west--east direction.

%%%%%%%%%%%%%%%%%%%%%%%%%%%%%%%%%%%%%%%%%%%%%%%%%%%%%%%%%%%%%%%%%%%%%%
We can now describe the tangent vector of a light ray $\lambda(s)$ by 
Boyer--Lindquist coordinates 
\begin{align}
	\dot{\lambda} &= \dot{r} \partial_{r} + \dot{\vartheta} \partial_{\vartheta}
	+ \dot{\varphi} \partial_{\varphi} + \dot{t} \partial_{t}
\label{eq:dotlambda1}
\intertext{and by celestial coordinates $\theta$ and $\psi$ for our 
moving observer}
	\dot{\lambda} &= \sigma \big( -\widetilde{e}_{0} 
		+ \sin\theta \cos\psi\; \widetilde{e}_{1} 
		+ \sin\theta \sin\psi\; \widetilde{e}_{2} 
		+ \cos\theta\; \widetilde{e}_{3} \big)
\label{eq:dotlambda2a}
\end{align}
where $\theta =0$ corresponds to the direction towards the black hole.
For the tetrad \eqref{eq:ObsTetradv} we observe the following dependencies
regarding \eqref{eq:ObsTetrad}:
\begin{equation}
\begin{alignedat}{3}
	&\widetilde{e}_{0} ={}& 
		k_{0r}\partial_{r} + k_{0\vartheta}\partial_{\vartheta} + 
		k_{0\varphi}\partial_{\varphi} + k_{0t}\partial_{t}, \\
	&\widetilde{e}_{1} ={}& 
		k_{1\vartheta}\partial_{\vartheta} + 
		k_{1\varphi}\partial_{\varphi} + k_{1t}\partial_{t}, \\
	&\widetilde{e}_{2} ={}& 
		k_{2\varphi}\partial_{\varphi} + k_{2t}\partial_{t}, \\
	&\widetilde{e}_{3} ={}&
		k_{3r}\partial_{r} + k_{3\vartheta}\partial_{\vartheta} + 
		k_{3\varphi}\partial_{\varphi} + k_{3t}\partial_{t}.
\end{alignedat}
\label{eq:ObsTetradDepend}
\end{equation}
Hence
\begin{equation}
\begin{aligned}
	\dot{\lambda}&= \sigma \bigl( 
		(-k_{0r} + k_{3r}\cos\theta)\partial_{r} 
		+ (-k_{0\vartheta} + k_{1\vartheta}\sin\theta \cos\psi
			+ k_{3\vartheta}\cos\theta )\partial_{\vartheta}  \\
	&\qquad + (-k_{0\varphi} + k_{1\varphi}\sin\theta \cos\psi 
			+ k_{2\varphi}\sin\theta \sin\psi 
			+ k_{3\varphi} \cos\theta)\partial_{\varphi}  \\
	&\qquad + (-k_{0t} + k_{1t}\sin\theta \cos\psi + k_{2t}\sin\theta \sin\psi
			+ k_{3t}\cos\theta)\partial_{t} \bigr).
\end{aligned}
\label{eq:dotlambda2b}
\end{equation}
Comparing coefficients of $\partial_{r}$, $\partial_{\vartheta}$, and  
$\partial_{\varphi}$ in \eqref{eq:dotlambda1} and \eqref{eq:dotlambda2b} 
yields
\begin{align}
	\dot{r} &= \sigma (-k_{0r} + k_{3r}\cos\theta), \\
	\dot{\vartheta} &= \sigma (-k_{0\vartheta} 
		+ k_{1\vartheta}\sin\theta \cos\psi + k_{3\vartheta}\cos\theta), \\
	\dot{\varphi} &= \sigma (-k_{0\varphi} + k_{1\varphi}\sin\theta \cos\psi 
		+ k_{2\varphi}\sin\theta \sin\psi + k_{3\varphi}\cos\theta).
\end{align}
These equations can be solved easily for $\cos\theta$ and $\sin\psi$
(using $k_{1\vartheta} \sin\theta\cos\psi = \tfrac{1}{\sigma}\dot{\vartheta} 
+ k_{0\vartheta} - k_{3\vartheta} \cos\theta$),
\begin{subequations}
\label{eq:thetapsi}
\begin{align}
	\cos\theta &= \frac{\frac{1}{\sigma}\dot{r} + k_{0r}}{k_{3r}}, 
	\label{eq:theta}
	\\
	\sin\psi &= \frac{ k_{3r}\bigl(\frac{1}{\sigma}\dot{\varphi} + k_{0\varphi} 
		- \frac{k_{1\varphi}}{k_{1\vartheta}}(
			\frac{1}{\sigma}\dot{\vartheta} +k_{0\vartheta})\bigr)
		- (k_{3\varphi}-\frac{k_{3\vartheta}}{k_{1\vartheta}}k_{1\varphi}) 
		(\frac{1}{\sigma}\dot{r} + k_{0r})}{
		k_{2\varphi}\sqrt{k_{3r}^{2}-(\frac{1}{\sigma}\dot{r} + k_{0r})^{2}} }
	\label{eq:psi}
\end{align}
\end{subequations}
where $\dot{\varphi}$, $\dot{\vartheta}$ and $\dot{r}$ have to be 
substituted from the equations of motion \eqref{eq:EoM_phi}, 
\eqref{eq:EoM_theta} and \eqref{eq:EoM_r}; since $\dot{r}$ and 
$\dot{\vartheta}$ are given as quadratic expressions, the signs have 
to be chosen consistently.
%$\dot{r}$ only positive sign, because we consider light rays reaching the observer.
%
%$\dot{\vartheta}$ positive sign: parametrization for upper part of boundary curve, 
%negative sign: parametrization for lower part of boundary curve

The remaining scalar factor $\sigma$ can be calculated analogously 
to $\alpha$ in \cite[Eq.~(20)]{GrenzebachPerlick.2014}. 
At first, express $\widetilde{e}_{0}$ \eqref{eq:ObsTetradv0} in terms of the 
tetrad $\{\partial_r, \partial_{\vartheta}, \partial_{\varphi}, \partial_t\}$
\begin{align}
	\widetilde{e}_{0} 
	&= \frac{1}{\sqrt{\Sigma}\sqrt{1-v^{2}}}
		\biggl( 
		\frac{(\Sigma + a \chi) \partial_t + a \partial_{\varphi}}{
		\sqrt{\Delta}}
		+ v_{1} \partial_{\vartheta}
		- v_{2}\frac{\partial_{\varphi} + \chi \partial_t}{\sin \vartheta} 
		- v_{3}\sqrt{\Delta} \, \partial_r
		 \biggr).
	\label{eq:tildee0}
\end{align}
As $\sigma = g\big(\dot{\lambda},\widetilde{e}_0 \big)$, 
see \eqref{eq:dotlambda2a}
we get $\sigma$ from \eqref{eq:Metric}, \eqref{eq:dotlambda1}, 
and \eqref{eq:tildee0},
\begin{equation}
\begin{split}
	\sigma = 
		\frac{1}{\sqrt{\Sigma}\sqrt{1-v^2}} 
		\biggl.\biggl(&
		\frac{a L_z - (\Sigma+a\chi)E}{\sqrt{\Delta}} \\
		&\quad + v_{1} \Sigma \dot{\vartheta}
		- v_{2} \frac{L_z - \chi E}{\sin\vartheta} 
		- v_{3} \frac{\Sigma}{\sqrt{\Delta}} \dot{r} 
		\biggr)\biggr\vert_{(r_O,\vartheta_O)} 
\label{eq:sigma}
\end{split}
\end{equation}
where $\dot{\varphi}$, $\dot{\vartheta}$ and $\dot{r}$ have to be substituted 
from \eqref{eq:EoM_phi}, \eqref{eq:EoM_theta} and \eqref{eq:EoM_r} as above. 

%Since $\sigma$ is negative, our lightrays are future directed parametrized.

With this expression, \eqref{eq:thetapsi} indeed describes the boundary curve 
of the black hole's shadow for a moving observer. The boundary represents
lightlike geodesics which, if you think of sending them from the observer's 
position into the past, reach the photon region asymptotically.
Each such geodesic must have constants of motion
\begin{align}
	K_{E} &= \left. \frac{4r^{2}\Delta}{(r-m)^{2}}\right|_{r=r_p}, &
	aL_{E} &= \left. \bigl(\Sigma + a\chi\bigr) - 
		\frac{2r\Delta}{r-m}\right|_{r=r_p},
\label{eq:LEKEc}
\end{align}
given by \eqref{eq:KL_SphLR} where $r_p$ is the radius coordinate of the 
limiting spherical lightlike geodesic. For $a>0$, this radius coordinate 
$r_p$ is---as in \cite{GrenzebachPerlick.2014}---extremal where the boundary 
of the exterior photon region intersects the cone $\vartheta = \vartheta_O$. 
Hence, the extremal values are the values of $r$ where \eqref{eq:regionK} 
holds with equality. 
Substituting $K_E$ and $L_E$ in \eqref{eq:thetapsi} and \eqref{eq:sigma}
by the expressions \eqref{eq:LEKEc} provides the shadow's
boundary curve $\big(\theta(r_p), \psi(r_p)\big)$ 
where $r_p$ runs between the extremal values. 

If $a=0$, then it is not possible to parametrize the boundary curve by 
$r_p$ because the right hand side of \eqref{eq:regionK} is zero, so 
it determines a unique $r_p$. By \eqref{eq:LEKEc} this results in an unique 
value for $K_E$ which, when inserted into \eqref{eq:thetapsi}, gives
the shadow's boundary curve in the form
$\big(\psi(L_E), \theta(L_E)\big)$. 
Here, $L_E$ ranges between the extremal values determined by 
\eqref{eq:EoM_theta} for $\Sigma^{2} \dot{\vartheta}^{2} = 0$.

%%%%%%%%%%%%%%%%%%%%%%%%%%%%%%%%%%%%%%%%%%%%%%%%%%%%%%%%%%%%%%%%%%%%%%
\section{Plots of Black Hole's Shadows}
\label{sec:Results}
\begin{figure}[bp]
\centering
	\includegraphics[scale=.71]{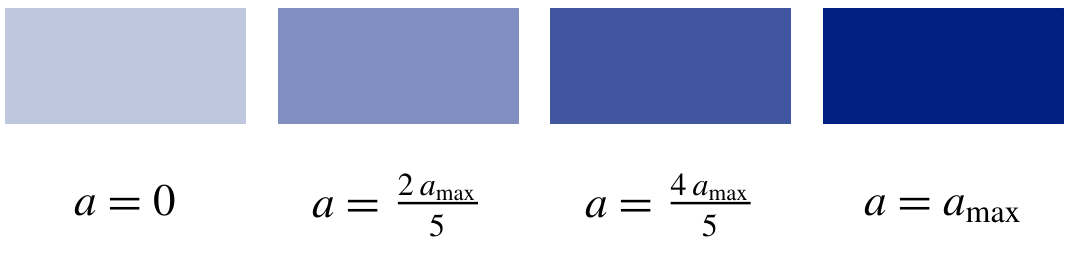}
	\caption{Legend for the different spins $a$ used for calculating 
	the black hole's shadows shown in Fig.~\ref{fig:Kerr}.}
	\label{fig:LegendSpin}
\end{figure}

\begin{figure}[tbp]
	\centering
	\includegraphics[scale=.71]{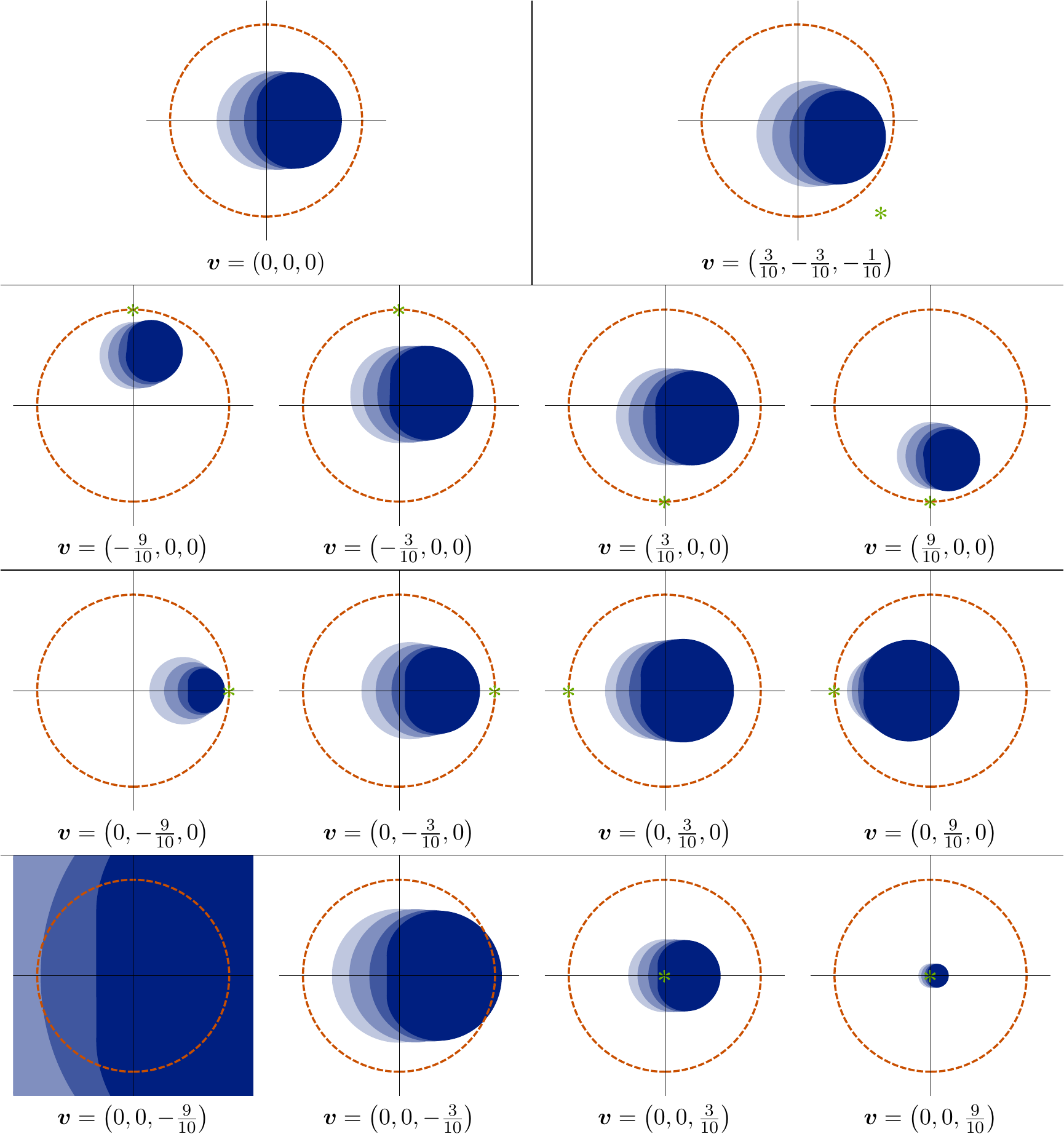}
	\caption{Aberrational effects on the shadows of Kerr black holes.
	The subfigures show the stereographic projection of the shadow for 
	observers moving with various velocities $\vec{v}=(v_{1},v_{2},v_{3})$ 
	which are noted beneath the plots ($r_{O}=5m$, $\vartheta_{O}=\frac{\pi}{2}$). 
	If $\vec{v}\neq 0$, the projected direction of the observer's motion 
	is marked by a (green) star. 
	Each plot combines the silhouettes for four different spins 
	($a=\kappa\cdot a_{\mathrm{max}}$ where $a_{\mathrm{max}}=m$), 
	see Fig.~\ref{fig:LegendSpin} for the corresponding legend.}
	\label{fig:Kerr}
\end{figure}

As described before, we used our analytical parameter representation 
\eqref{eq:thetapsi} with \eqref{eq:sigma} and \eqref{eq:LEKEc} to 
calculate the boundary curve of the shadow as seen by an observer 
moving with four-velocity $\widetilde{e}_0$.
The results in Figs.~\ref{fig:Kerr} are visualized via stereographic 
projection from the celestial sphere onto a plane, as illustrated in 
Fig.~\ref{fig:Beob}. Standard Cartesian coordinates in this plane are 
given by
\begin{equation}
\begin{aligned}
	x(r_p) &= -2 \tan \Big( \frac{\theta(r_p)}{2} \Big)
		\sin \big( \psi(r_p) \big),\\
	y(r_p) &= -2 \tan \Big( \frac{\theta(r_p)}{2} \Big)
		\cos \big( \psi(r_p) \big).
\end{aligned}
\label{eq:stereo}
\end{equation}
All plots of shadows shown in Fig.~\ref{fig:Kerr} belong to Kerr black 
holes where each subfigure combines the pictures for four spin values, 
see legend in Fig.~\ref{fig:LegendSpin}.

In principle, the shadows for moving observers ($\vec{v}\neq 0$) 
are calculable from the shadow seen by the standard observer ($\vec{v}=0$)
with the help of Penrose's aberration formula \cite{Penrose.1959}
\begin{equation}
	\tan \frac{\widetilde{\alpha}}{2} = \sqrt{\frac{c-v}{c+v}}
	\tan \frac{\alpha}{2}.
\label{eq:aberration}
\end{equation}
But for applying this formula, one may need to make coordinate transformations
since the angles $\alpha$ and $\widetilde{\alpha}$ have to be measured 
against the direction of the motion.
Hence, no transformations are needed if the observer moves in radial 
direction. Then, the shadow is magnified if the observer moves away from 
the black hole, and demagnified if the observer moves towards the black hole.
In this case, our formula \eqref{eq:theta} reduces to the following common 
variant of Penrose's aberration formula \eqref{eq:aberration}
\begin{equation}
	\cos \widetilde{\theta} = \frac{v+\cos\theta}{1+v\cos\theta}.
\label{eq:aberrationCos}
\end{equation}

Penrose emphasized in his article \cite{Penrose.1959} that the aberration
formula maps circles on the celestial sphere onto circles. 
%Penrose:
% ÒIt turns out, in particular, that the appearance of a sphere, no matter how 
% it is moving, is always such as to present a circular outline to any observer. 
% Thus an instantaneous photograph* of a rapidly moving sphere has the same 
% outline as that of a stationary sphere.Ó
Thus, the shadow of a non-rotating black hole ($a=0$) is always circular, 
independent of the observer's motion. Consequently, our pictures of the shadow
are then always circular, because the stereographic projection 
\eqref{eq:stereo} maps circles onto circles, too.

%%% Kerr
Fig.~\ref{fig:Kerr} shows several pictures of shadows for differently 
moving observers in Kerr space-times. The results for the standard 
observer ($\vec{v}=0$, $\widetilde{e}_\mu = e_\mu$) are shown in the 
left plot in the first row of Fig.~\ref{fig:Kerr}. The right plot
is exemplary and belongs to an observer moving with velocity 
$\vec{v}=\bigl(\tfrac{3}{10},-\tfrac{3}{10},-\tfrac{1}{10}\bigr)$.
In each of the lower rows we vary only one component $v_i$ of $\vec{v}$; 
in the following we write $\vec{v_i}$ as abbreviation for those observer 
velocities $\vec{v}$ with $v_i\neq 0$ and $v_{j}=0$, $j\neq i$.
Due to our definition of the tetrad $e_\mu$ in \eqref{eq:ObsTetrad} and 
of the observer's four-velocity $\widetilde{e}_0$ in \eqref{eq:ObsTetradv0}
the observer moves in $\vartheta$ direction if $\vec{v}=\vec{v_1}$, and in 
$r$ i.e. radial direction if $\vec{v}=\vec{v_3}$. For $\vec{v}=\vec{v_2}$ 
the motion is in $\varphi$ direction.

%4th row: varying $v_1$, $v_2=v_3=0$ radial motion
Since the last row of Fig.~\ref{fig:Kerr} shows the plots of shadows 
seen by a radially moving observer ($\vec{v}=\vec{v_3}$), the shadows are 
magnified if the observer moves away from the black hole ($v_3$ negative), 
and demagnified, if the observer moves towards the black hole 
($v_3$ positive), as mentioned before.

%2nd row: varying $v_1$, $v_2=v_3=0$
%3rd row: varying $v_2$, $v_1=v_3=0$
For velocities $\vec{v}=\vec{v_1}$ or $\vec{v}=\vec{v_2}$ the shadow is 
shifted in the direction of the observer's motion with bigger effects 
for higher velocities. Also the size of the shadow is affected. But all
these aberrational changes are explainable if one relates the direction
of the observer's motion to the spin of the black hole and to the 
equatorial plane as symmetry plane. 

Furthermore, the shadow is symmetric with respect to a horizontal 
axis as long as the observer does not move in $\vartheta$ direction 
because $\sin\psi$, see \eqref{eq:psi}, depends on $\dot{\vartheta}$
which is given by a quadratic expression, see \eqref{eq:EoM_theta}.
%\eqref{eq:psi}, \eqref{eq:ObsTetradvi}, \eqref{eq:ObsTetradDepend},
Hence, the different signs of $\dot{\vartheta}$ yield different solutions of 
\eqref{eq:thetapsi} for the points $(\theta,\psi)$ and $(\theta,\pi-\psi)$.
Without a $\vartheta$ component in the velocity, the symmetry of 
the shadow is not affected even if the observer is not in the equatorial 
plane, i.e. $\vartheta_O\neq \tfrac{\pi}{2}$. 

%$\vec{v}=0$
%\begin{align}
%	\cos\theta &= \frac{\dot{r}}{\sigma k_{3r}}, &
%	\sin\psi &= \frac{\dot{\varphi} + \sigma k_{0\varphi}}{
%		\sigma k_{2\varphi} \sin\theta}
%	\label{eq:thetapsiV0}
%\end{align}

All in all, the shadows shown in Fig.~\ref{fig:Kerr} are calculated 
for relatively fast moving observers ($v=0.3\,c$ up to $v=0.9\,c$). 
Thus, the aberrational influence for the future observations of the
shadow of Sgr~A* within the Event Horizon Telescope or the Black Hole 
Cam project is expected to be very small since our solar system 
orbits the galactic center with roughly 
$250\unit{\frac{km}{s}} \approx \frac{1}{1000}c$, see \cite{ReidMenten.2009}.
Nevertheless, the study of aberrational effects are of interest
from a fundamental point of view.

%Since our solar system orbits the galactic center with roughly 
%$250\unit{\frac{km}{s}} \approx \frac{1}{1000}c$, see \cite{ReidMenten.2009},
%the effects of the observer's motion on the shadow are very small.
%circular rotation speed $\Theta_0 = 254 \pm 16\unit{\frac{km}{s}}$.
%\cite{ReidMenten.2009}

%%%%%%%%%%%%%%%%%%%%%%%%%%%%%%%%%%%%%%%%%%%%%%%%%%%%%%%%%%%%%%%%%%%%%%
\section*{Acknowledgments}
I would like to thank Volker Perlick, Claus L\"ammerzahl, Nico Giulini, 
Norman G\"urlebeck, Eva Hackmann, Valeria Diemer (n\'{e}e Kagramanova), 
Jutta Kunz and Luciano Rezzolla for helpful discussions.
Furthermore, I want to thank Dirk P\"utzfeld, Claus L\"ammerzahl, and 
Bernard F. Schutz for organizing the 524th WE-Heraeus-Seminar and 
for the opportunity to contribute to this proceedings volume. 
%My gratitudes for offering poster awards deserves the WE-Heraeus foundation. 
The WE-Heraeus foundation deserves my gratitudes for offering 
poster awards. 
I gratefully acknowledge support from the DFG within the Research 
Training Group 1620 \blqq Models of Gravity\brqq\ and from the 
\blqq Centre for Quantum Engineering and Space-Time Research (QUEST)\brqq.

%%%%%%%%%%%%%%%%%%%%%%%%%%%%%%%%%%%%%%%%%%%%%%%%%%%%%%%%%%%%%%%%%%%%%%
\bibliographystyle{unsrt}

%\bibliography{grenzebach_eom_proceedings_2013}

\end{document}